# P-adic arithmetic coding.

*Anatoly Rodionov, Sergey Volkov*
*Spectrum Systems, Inc*

## Abstract

A new incremental algorithm for data compression is presented. For a sequence of input symbols algorithm incrementally constructs a p-adic integer number as an output. Decoding process starts with less significant part of a p-adic integer and incrementally reconstructs a sequence of input symbols. Algorithm is based on certain features of p-adic numbers and p-adic norm.

p-adic coding algorithm may be considered as of generalization a popular compression technique – arithmetic coding algorithms. It is shown that for p = 2 the algorithm works as integer variant of arithmetic coding; for a special class of models it gives exactly the same codes as Huffman's algorithm, for another special model and a specific alphabet it gives Golomb-Rice codes.

> *"For more than forty years I've been speaking in prose without even knowing it!"*
>
> Moliere. *Le Bourgeois Gentilhomme.*

## Introduction

Arithmetic coding algorithm in its modern version was published in *Communications of ACM* in June 1987 [**Witten**], but the authors, Ian Witten, Radford Neal and John Cleary, referred to [**Abrahamson**] as to "the first reference to what was to become the method of arithmetic coding". So we may say that it is known *"for more than forty years"*. The algorithm now is a common knowledge – it was published in numerous textbooks (see for example [**Salomon, Sayood**]), some reviews were published [**Bodden**, **Said**], *Dr. Dobb's Journal* popularized it [**Nelson**], wiki [**wiki**] contains an article about it, a lot of sources could be found on web... So why one more paper on this subject and what is this "p-adic arithmetic"?

Let go back to the original idea of arithmetic coding. In arithmetic coding a message is represented as a subinterval [b, e) of union semi interval [0, 1). (We will give all definitions later) When a new symbol s comes a new subinterval [b(s), e(s)) of [b, e) is constructed. Common method to calculate a new subinterval is to divide a current interval into |A| (A is an alphabet, |A|- number of symbols) subintervals, each subinterval represents a symbol from A and has length proportional to probability of this symbol. For a new symbol s corresponding subinterval [b(s), e(s)) will be return by encoder. Thus encoding is a process of narrowing intervals (we will call them message intervals) starting from the union interval:

[0, 1) ≡ [$b_0$, $e_0$), [$b_1$, $e_1$), [$b_2$, $e_2$), … , [$b_t$, $e_t$)
where
$0 = b_0, \leq b_1 \leq b_2 \leq … \leq b_t$
$1 = e_0, \geq e_1 \geq e_2 \geq … \geq e_t$
All $b_i$ and $e_i$ are real numbers.

A last constructed subinterval may be used as a final output, or any point x from last subinterval and message length. But usually a special symbol EOM (*End Of Message*), which does not belong to the alphabet, is used as termination symbol of a message. In this case only a point x can be used as coding result.

Decoding is also a process on narrowing intervals. It starts with union interval and a point x inside it. Decoder finds a symbol by dividing current intervals into |A| subintervals and finds the one that contains point x, say [$b_1$, $e_1$). Corresponding to this interval symbol $s_1$ is pushed into an output buffer; [$b_1$, $e_1$) is used as a new current interval. And so on until EOM symbol is received.

But here is a problem – one have to use infinite precision real numbers to implement this algorithm and there is no such a thing like effective infinite precision real arithmetic. This problem was always considered as a technical one. Solution is simple - just use integers instead. There is a canonical implementation, first written in *C* [**Witten**], which was later reproduced in other languages, but no analysis of what happens to the algorithm after moving it from the field of real numbers to the ring of integer numbers was published.

In this paper we introduce p-adic arithmetic coding which is based on mapping a message to a path on a p-tree (a tree with p outgoing branches at each vertex; we also assume that p is a prime number). This path is constructed as a common part of paths to the left and right edges of a subinterval $[g_l(s), g_r(s))$, where $g_l(s)$ and $g_r(s)$ are from a special equidistant grid G on [0, 1). This semi interval is constructed according to the same rules, as in real number arithmetic coding, but in contrast to it, the edges are not arbitrary real numbers, but belong to the grid.

A path on a p-tree can be naturally presented as a p-adic integer number. p-adic distance proved to be a natural measure on paths – the longer a common part of two paths, the smaller p-adic distance between them. Function $ord_P$, also known as p-adic logarithm, gives length of a common path.

A path can be also identified by its final point on a grid. A grid point g can be represented by an integer index k from a finite integer ring as $g=k*|G|^{-1}$ (here $|G|$ is number of elements in G). The crucial point of this algorithm is how we can calculate a path from an index and vice versa – an index from a path. *IP* (*Index-Path*) mapping, described in this article, presents an elegant and efficient way for this.

Now we ready to give a brief sketch of how does the algorithm work. As initial step we have to define an alphabet A, a model M, an output buffer (it will contain a p-adic integer number B) and a grid G on [0, 1). Start with union coding semi interval represented by two indexes l=0 (left) and r=0 (right), and B=0. When a new symbol s comes, the model calculates a new subinterval [l(s), r(s)) (l and r are indexes from a finite integer ring, while l^ and r^ – paths presented as p-adic integer numbers). Using *IP* transformation (we use symbol ^ for this transformation) we can calculate p-adic representation of paths to these edges l(s)^ and r(s)^. If p-adic distance between them is equal to 1, we continue encoding using [l(s), r(s)) as a new current encoding interval. If the distance is less then 1, then l(s)^ and r(s)^ have a common path of length $c=ord_P(l(s)^\wedge, r(s)^\wedge)$. That means that path to any point inside [l(s), r(s)) have the same first (least significant) c digits as l(s)^. We can push this common path to an output buffer adding them as new most significant part of p-adic number B. We can also drop c least significant digit of l(s)^ and r(s)^.

Both of these operations are possible, because p-adic numbers are read from left to right, i.e. less significant digit (those that are multiplied by less powers of p) are in the left part of buffer. This feature of p-adic integer numbers explains why p-adic arithmetic coding and decoding are incremental.

Now we can continue encoding with truncated l(s)^ and r(s)^. To do this we must calculate new subinterval, corresponding to new paths. This also can be done using *IP* transformation. Encoding will continue using [(l(s)^)^, (r(s)^)^) as new current message interval from some grid. This procedure we will call *PR* rescaling. In the case of p=2 this is procedure is similar to well known *E1/E2* rescaling [**Bodden**]. But *PR* rescaling gives a better insight of this mechanism, connects it with p-adic norm and can be used for any prime p. Moreover, *PR* rescaling is more accurate on boundaries and because of this the algorithm is able to reproduce Huffman codes for certain models. We will also generalize *E3* rescaling, which is based on usual Archimedean norm (absolute value in this case), for any prime p.

p-adic arithmetic coding algorithm generalizes not only arithmetic coding. For a special class of models, p-adic coding algorithm works exactly as Huffman's algorithm [**Huffman**]. In this models weights of all symbols should be equal to $p^{-n}$, where n some positive numbers and a sum of all weight is equal to 1. In other words, they are leaves of a Huffman code tree.

For a special model and one symbol alphabet p-adic arithmetic coding reproduce Golomb-Rice codes [**Golomb**, **Rice**].

# *Definitions*

## Alphabet

*Alphabet* A - A non empty set of symbols $a_i$. |A| - number of symbols in A.

In most examples below 4 symbols alphabet [a, b, c, d] will be used. Other examples: binary alphabet [0, 1], 128 characters ASCII, alphabet of 256 different eight-bit characters. The last one is used in all tests. Even an alphabet containing only one symbol makes sense – as it will be shown, the algorithm creates exactly Golomb-Rice [**Golomb, Rice**] codes in this case.

## Message

*Message M* – a sequence of symbols from alphabet A.

$M = (a_0, a_1, \ldots, a_i, \ldots, a_n)$

where $a_i$ belong to A.
Example: (a, b, a, a, b, c, d, a).

## Semi interval

[l, r) – includes l, but not r.

Notation [,) means that the left point is included to the interval, while the right one is not.
Below we always deal with subintervals of [0, 1).

## Grid

Later we will subdivide [0, 1) into $P^N$ (P and N are natural numbers) semi intervals of equal length. Each has length $P^{-N}$ and can be identified by its left edge. These left edge points form a grid $G(P^N)$. Coordinate of a point of the grid with index k is evidently $kP^{-N}$. We will use notation $g_k(N)$ for points from $G(P^N)$.

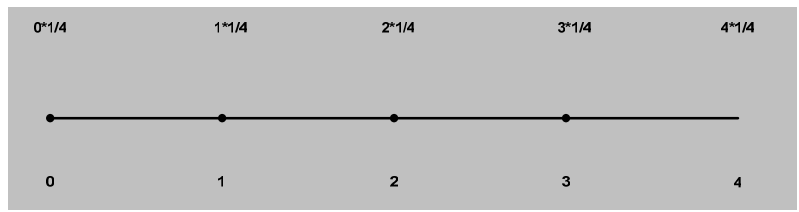

*Picture 1. Grid $P^2$ (P=2)*

These indexes will play an important role in our discussion. If fact, all calculations will be done using indexes. The range of indexes is

$0 \leq k < P^N$

In other words, indexes are nonnegative numbers modulo $P^N$. Negative numbers are defined in this ring as

$-k = P^N - k$

## Weight interval

Following the main idea of arithmetic coding let map alphabet A to a semi interval [0, 1), which we will refer as a *weight interval*. To do this enumerate symbols from A in any order (the order is not important in a sense that compression rate does not depend on it) and divide the interval in |A| semi intervals. Semi interval $[w_i, w_{i+1})$ corresponds to symbol $a_i$. To make compression effective lengths of these intervals must be equal to probability of symbols in a message:

$| w_i - w_{i+1} | = p_i$

where $p_i$ – probability of symbol $a_i$.

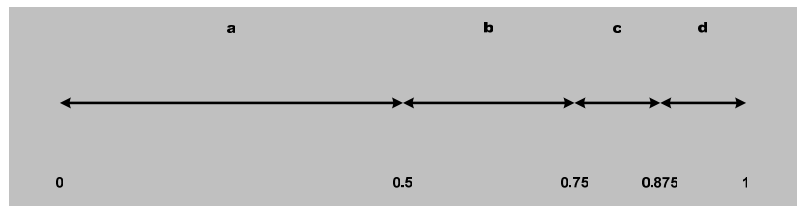

*Picture 2. Weight interval*

To define weight intervals we will use notation:

{ symbol$_0$:[semininterval$_1$), … , symbol$_{|A|-1}$:[semininterval$_{|A|-1}$) }

For example:

{a:[0, 0.5), b:[0.5, 0.75), c:[0.75, 0.875), d:[0.875,1)}

## Message interval

Let fix a natural number N, a prime number P and create a grid $G(P^N)$. Messages will be mapped to semi intervals [l, r) of this interval.

0 ≤ l < r < 1

l, r belongs to $G(P^N)$.

Arithmetic coding is just a process of narrowing a message interval. When a new symbol comes, a current message interval is divided in |A| subintervals proportional to weight interval and then a subinterval corresponding to a new symbol is selected as a new message interval. Thus starting with [0, 1) (empty message) interval we end up with a subinterval corresponding to the whole message.

For example let see how a short message {a, b, a} may be coded (here we use weight interval from previous section):

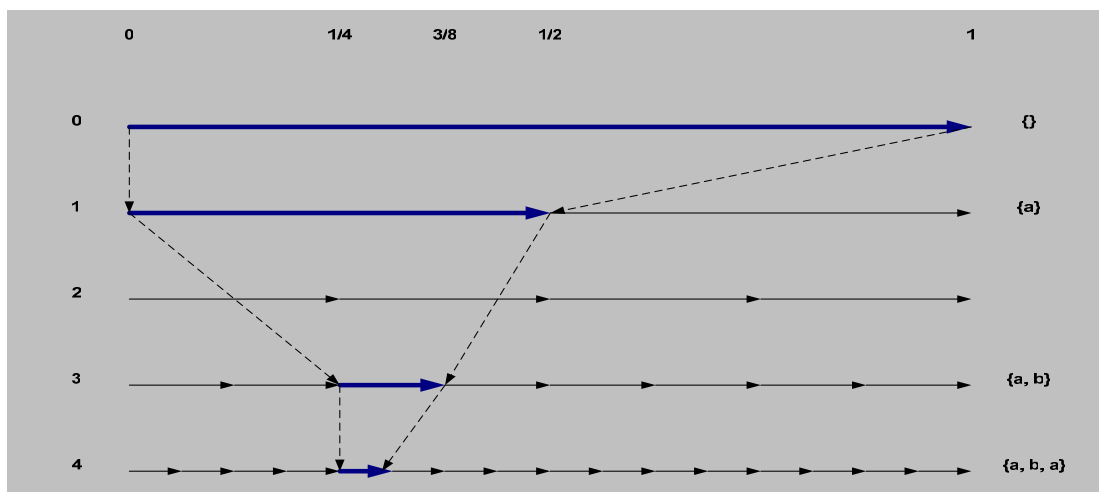

*Picture 3. Message interval*

We may also present this in a table (using actual values of $g_k(N)$):

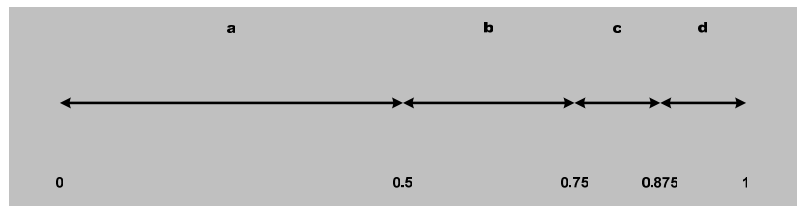

*Picture 2. Weight interval*

To define weight intervals we will use notation:

{ symbol$_0$:[semininterval$_1$), … , symbol$_{|A|-1}$:[semininterval$_{|A|-1}$) }

For example:

{a:[0, 0.5), b:[0.5, 0.75), c:[0.75, 0.875), d:[0.875,1)}

## Message interval

Let fix a natural number N, a prime number P and create a grid $G(P^N)$. Messages will be mapped to semi intervals [l, r) of this interval.

0 ≤ l < r < 1

l, r belongs to $G(P^N)$.

Arithmetic coding is just a process of narrowing a message interval. When a new symbol comes, a current message interval is divided in |A| subintervals proportional to weight interval and then a subinterval corresponding to a new symbol is selected as a new message interval. Thus starting with [0, 1) (empty message) interval we end up with a subinterval corresponding to the whole message.

For example let see how a short message {a, b, a} may be coded (here we use weight interval from previous section):

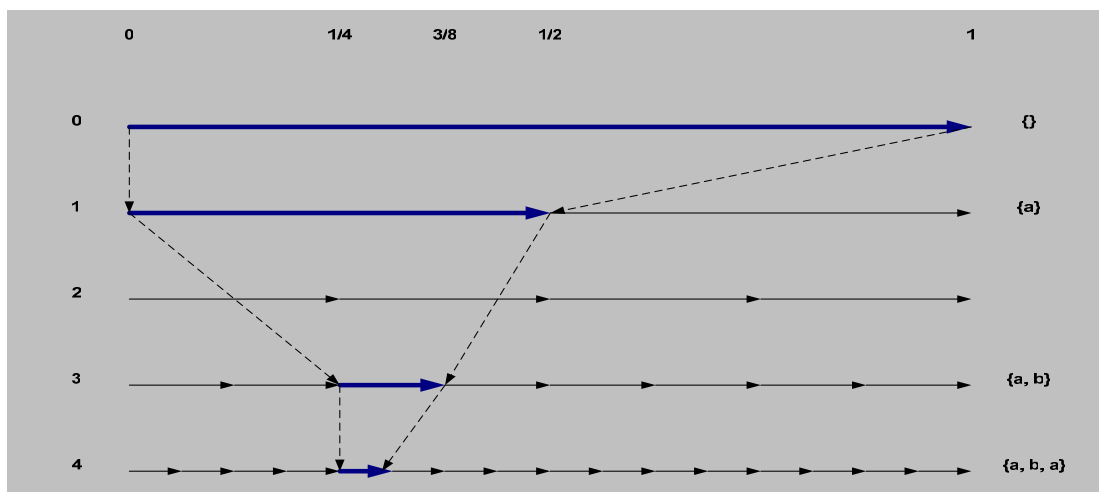

*Picture 3. Message interval*

We may also present this in a table (using actual values of $g_k(N)$):

| Message | Semi interval | Numerical value |
|---|---|---|
| { } | [ $g_0(0)$ , $g_0(0)$ ) | [ 0 , 1 ) |
| {a} | [ $g_0(1)$ , $g_1(1)$ ) | [ 0 , 1/2 ) |
| {a, b} | [ $g_2(3)$ , $g_3(3)$ ) | [ 2/4 , 3/8 ) |
| {a, b, a} | [ $g_4(4)$ , $g_5(4)$ ) | [ 4/8 , 5/8 ) |

An important difference from the original idea of real number arithmetic coding is that here we use only points from a grid as subintervals edges.

## *Coding tree*

Consider a tower of grids

$G(P^0) < G(P^1) < G(P^2) < \ldots < G(P^n) \ldots$

By construction, if a point belongs to $G(P^n)$, then it belongs to $G(P^{n+1})$, $G(P^{n+2})$,…, $G(P^k)$ (k>n). $G(P^0)$ consist only of one point.

Now let construct a coding tree. Start with a root – which is evidently the only point from $G(P^0)$ - $g_0(0)$ . Then comes a first level – points from $G(P^1)$. Link the root $g_0(0)$ with points from $G(P^1)$: $g_0(1)$, $g_1(1)$, … , $g_{P-1}(1)$. This gives us first level of the coding tree. Now we can continue. Let us assume that the tree build up to a level N. To create a new N+1 level, we have to:
    Construct a new grid $G(P^{n+1})$ as a new bottom level to the bottom (us usual, tree grows downwards).
    Link points from the last level (i.e. points from $G(P^n)$) to points from $G(P^{n+1})$ according to the following rule:

$g_k(n)$ link to points $g_{k*P}(n+1)$, $g_{(k+1)*P}(n+1)$, …, $g_{(k+P-1)*P}$ (n+1)

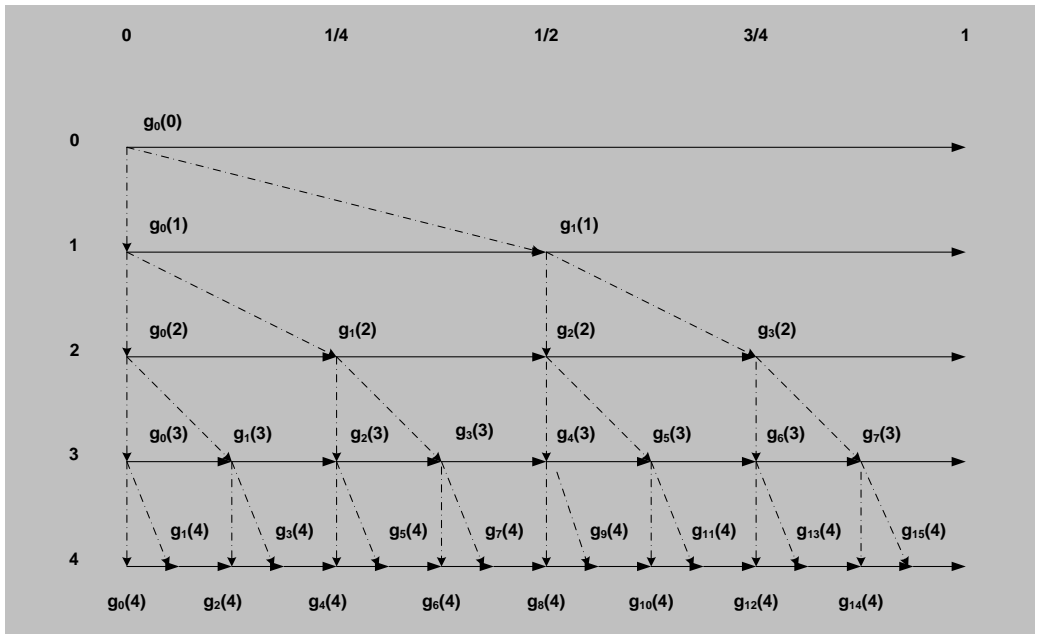

*Picture 4. Coding tree*

Here, as in most other illustrations, we use P=2 to simplify drawing.

Now we can use the grid and the tree to code a simple message {a, b, a} using weights {a:[0, 1/2), b:[1/2, 3/4), c:[3/4, 7/8), d:[ 7/8, 1)}

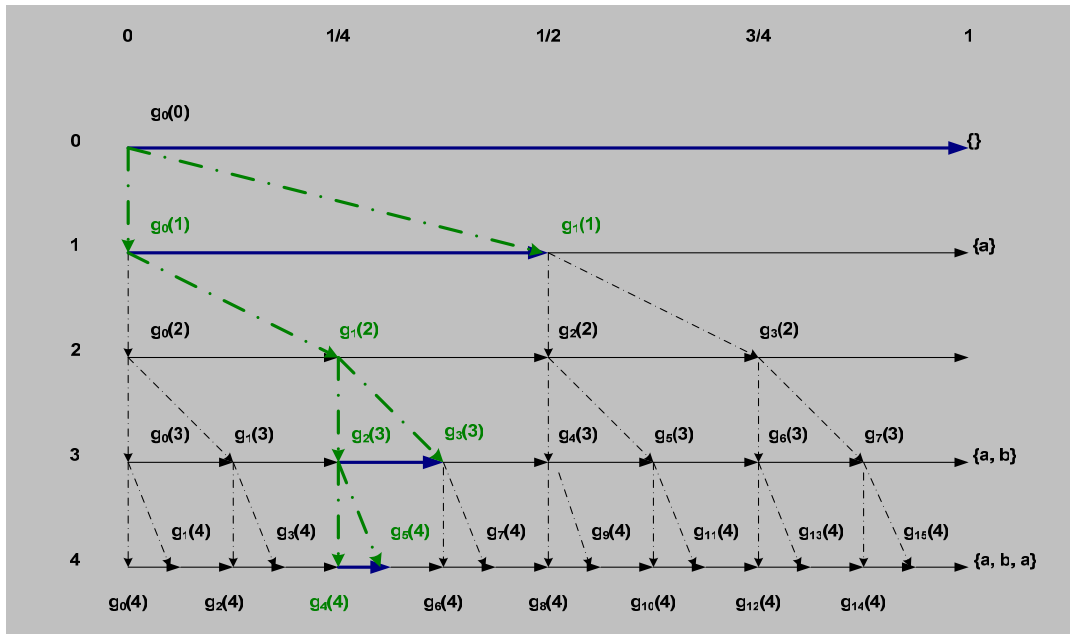

*Picture 5. Paths on coding tree*

| Message | Semi interval | Path |
|---|---|---|
| { } | [ $g_0(0)$ , $g_0(0)$ ) | [ { $g_0(0)$ }, { $g_0(0)$ } ) |
| {a} | [ $g_0(1)$ , $g_1(1)$ ) | [ { $g_0(0)$ , $g_0(1)$ } , { $g_0(0)$ , $g_1(1)$ } ) |
| {a, b} | [ $g_2(3)$ , $g_3(3)$ ) | [ { $g_0(0)$, $g_0(1)$ , $g_1(2)$ , $g_2(3)$ }, { $g_0(0)$, $g_0(1)$ , $g_1(2)$ , $g_3(3)$ } ) |
| {a, b, a} | [ $g_4(4)$ , $g_5(4)$ ) | [ { $g_0(0)$, $g_0(1)$ ,$g_1(2)$ ,$g_2(3)$,$g_4(4)$}, { $g_0(0)$,$g_0(1)$ ,$g_1(2)$ ,$g_2(3)$,$g_5(4)$ } ) |

We need a more convenient way to refer to grid points and tree paths. Grid points can be easily represented as indexes, i.e. well known positive integer numbers, while for paths we will use p-adic integer numbers.

### Representation of paths as p-adic integer numbers

From any point $g_k(n)$ we have P different links to a next (n+1) level. We can mark our next choice with a nonnegative integer number m

$0 \le m_j < P$
$0 \le j \le n$

Now we can represent any (final) paths on the coding tree as a vector

$M = \{m_0, m_1, m_2, \ldots , m_n\}$

This vector can be mapped to a nonnegative number x

$x = m_0 P^0 + m_1 P^1 + m_2 P^2 + \ldots + m_n P^n$

This mapping is evidently one to one. The number x may be considered as a p-adic integer number. These numbers are not well known among programmers. One can find an introduction in p-adic numbers in [**Baker, Koblitz**]. A very helpful way to visualize some unusual properties of p-adic mathematic may be found in [**Holly**].

The first coefficient $m_0$ tells us to what top level subinterval of [0, 1) the point belongs. The next one $m_1$ – to which subinterval of this interval the point belongs, and so on.

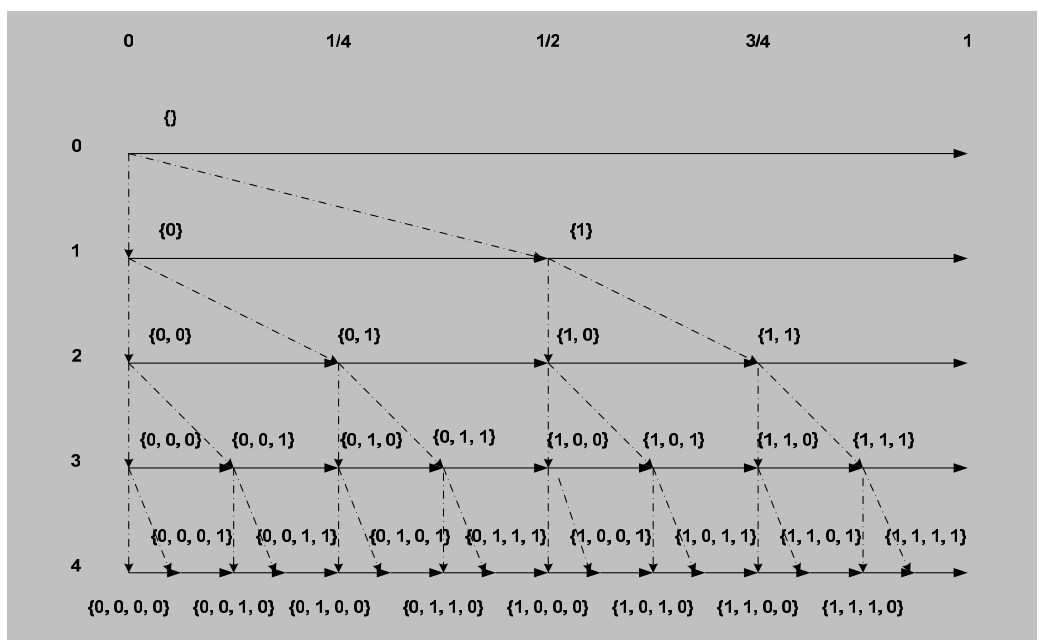

*Picture 6. p-adic representation of binary tree.*

| Level | Paths | | | | | | | | |
|---|---|---|---|---|---|---|---|---|---|
| 0 | {} | | | | | | | | |
| 1 | Paths | {0} | {1} | | | | | | |
|   | p-adic number | 0 | 1 | | | | | | |
| 2 | Paths | {0,0} | {0,1} | {1,0} | {1,1} | | | | |
|   | p-adic number | 0 | 2 | 1 | 3 | | | | |
| 3 | Path | {0,0,0} | {0,0,1} | {0,1,0} | {0,1,1} | {1,0,0} | {1,0,1} | {1,1,0} | {1,1,1} |
|   | p-adic number | 0 | 4 | 2 | 6 | 1 | 5 | 3 | 7 |

A tree for P=3 is shown in the next picture

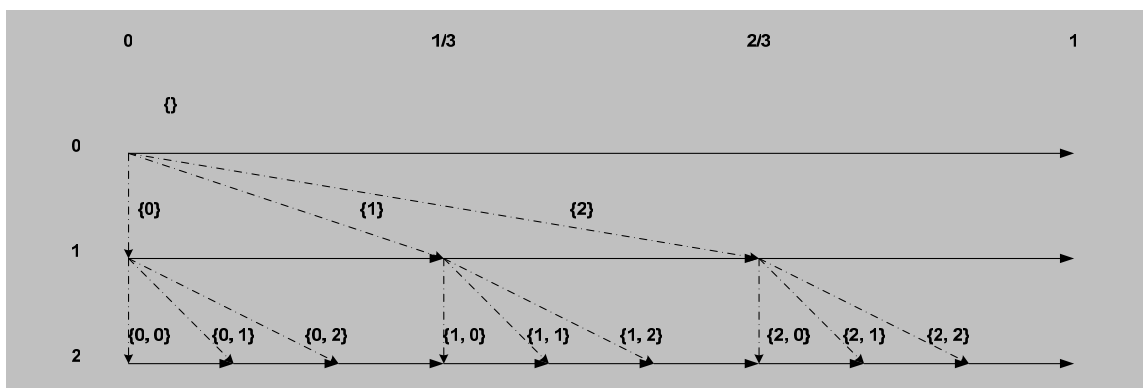

*Picture 6. p-adic representation of a tree; P=3.*

| Level | Paths |  |  |  |  |  |  |  |  |
|---|---|---|---|---|---|---|---|---|---|
| 0 | {} |  |  |  |  |  |  |  |  |
| 1 | Paths | {0} | {1} | {2} |  |  |  |  |  |
|   | p-adic number | 0 | 1 | 2 |  |  |  |  |  |
| 2 | Paths | {0,0} | {0,1} | {0,2} | {1,0} | {1,1} | {1,2} | {2,0} | {2,1} | {2,2} |
|   | p-adic number | 0 | 3 | 6 | 1 | 4 | 7 | 2 | 5 | 8 |

In our algorithm we will use p-adic distance. This distance can be defined with the help of p-adic logarithm function, usually called as $ord_P$.

$ord_P(x)$ = max number r such that $(x \% P^r) = 0$; $x \neq 0$

p-adic norm is

$|x|_P = 1 / P^{**}ord_P(x)$     $x \neq 0$

$|0| = 0$.

and distance

$d_P(x,y) = |x - y|_P$

It can be shown [**Koblitz**] these are "*real*" norm and distance, i.e. all three axioms are valid for them.

For two paths x and y

$x = m_0P^0 + m_1P^1 + m_2P^2 + \ldots + m_nP^n$

$y = k_0P^0 + k_1P^1 + k_2P^2 + \ldots + k_nP^n$

$ord_P(x-y)$ gives a number of common links. $d_P(x,y)$ have a very intuitive meaning – the greater number of common links have two paths, the closer they are in terms of p-adic distance.

## Index representation

Now return to the first method of mapping paths – by end points. An end point belongs to a grid $G(P^n)$, so it is defined by its index **a** which is just a plain nonnegative integer number. How this index is connected to the paths leading to this point?
Let x be a path

$x = m_0P^0 + m_1P^1 + m_2P^2 + \ldots + m_nP^n$

$0 \leq m_j < P$
$0 \leq j \leq n$

Then the path x ends at a point g at level n

$g = m_0P^{-1} + m_1P^{-2} + m_2P^{-3} + \ldots + m_nP^{-n-1}$

We just negate powers and subtract 1. It can be proved by simple induction.
Let **a** be an index corresponding to point g on the grid.

$g = aP^{-n-1}$

$a = m_0P^n + m_1P^{n-1} + m_2P^{n-2} + \ldots + m_nP^0$

We can rewrite **a** in usual form

$a = u_0P^0 + u_1P^1 + u_2P^2 + \ldots + u_nP^n$

If we consider **x** and **a** as integer numbers then mapping is just reversing vectors of coefficients. This mapping we will call **IP** (*Index-Path*) mapping.

We will introduce some useful notations in the next section and use this feature intensively in algorithms: we can perform ordinary integer arithmetic operations on indexes calculating new subinterval and immediately get paths to them.

Mapping paths to points was considered in a more general form by S.V. Kozyrev [**Kozyrev**]. Following his notation we will use symbol ρ for it. The ρ mapping has a very important feature

$|\rho(x) - \rho(y)| \leq |x - y|_P$

were $|\rho(x) - \rho(y)|$ is Archimedean (our usual) distance which in our case is absolute value. This means that if two paths are close to each other then corresponding end points are also close in our usual Archimedean norm. A proof of this can be found in [**Kozyrev**].

ρ mapping is not one to one, ***IP*** (Index-Path), mapping which deals with finite sums, is one to one mapping. Let, as previously a and g be

$a = m_0 P^n + m_1 P^{n-1} + m_2 P^{n-2} + \ldots + m_n P^0$

$g = a P^{-n-1}$

To what subinterval of level 1 of [0, 1) the point belongs to? It depends on the value of $m_0$ only. Straightforward calculations

$m_1 P^{n-1} + m_2 P^{n-2} + \ldots + m_n P^0 \leq (P-1)(P^0 + P^1 + P^2 + \ldots + P^{n-1}) =$
$(P-1)((P^n - 1)/(P-1)) = (P^n - 1)$

show that a sum of all less significant terms can't move a point to another subinterval. With $m_0$ fixed, we can conclude that $m_1$ solely defines subinterval inside the second subinterval and so on.
Let us continue with the example (with P=2) above by adding indexes of grid points to the table:

| Level | Paths | | | | | | | | |
|---|---|---|---|---|---|---|---|---|---|
| 0 | {} | | | | | | | | |
| 1 | Paths | {0} | {1} | | | | | | |
|  | p-adic number | 0 | 1 | | | | | | |
|  | Index | 0 | 1 | | | | | | |
|  | Points | 0 | 1/2 | | | | | | |
| 2 | Paths | {0,0} | {0,1} | {1,0} | {1,1} | | | | |
|  | p-adic number | 0 | 2 | 1 | 3 | | | | |
|  | Index | 0 | 1 | 2 | 3 | | | | |
|  | Points | 0 | 1/4 | 2/4 | 3/4 | | | | |
| 3 | Paths | {0,0,0} | {0,0,1} | {0,1,0} | {0,1,1} | {1,0,0} | {1,0,1} | {1,1,0} | {1,1,1} |
|  | p-adic number | 0 | 4 | 2 | 6 | 1 | 5 | 3 | 7 |
|  | Index | 0 | 1 | 2 | 3 | 4 | 5 | 6 | 7 |
|  | Points | 0 | 1/8 | 2/8 | 3/8 | 4/8 | 5/8 | 6/8 | 7/8 |

As an illustration let us compare Archimedean and p-adic distances for the following three points $g_2(3)$ (or {0, 1, 0}), $g_3(3)$ ({0,1,1}) and $g_4(3)$ ({1, 0, 0}). Archimedean distance we all used to:

| Points (p-adic) | 2 | 3 | 4 |
|---|---|---|---|
| 2 | 0 | 1/8 | 1/4 |
| 3 | 1/8 | 0 | 1/8 |
| 4 | 1/4 | 1/8 | 0 |

p-adic logarithm (ord) has values:

| Points (p-adic) | 2 | 6 | 1 |
|---|---|---|---|
| 2 |   | 2 | 0 |
| 6 | 2 |   | 0 |
| 1 | 0 | 0 |   |

and **p**-adic distances are:

| Points (p-adic) | 2 | 6 | 1 |
|---|---|---|---|
| 2 | 0 | 1/4 | 1 |
| 6 | 1/4 | 0 | 1 |
| 1 | 1 | 1 | 0 |

From this we may observe that the greater is a common path, the closer are points in **p**-adic norm.

## Operators ^ and []

Let us define operator **^** to transform points from index representation to path representation, or, in other words, from nonnegative integers modulo $P^N$ to **p**-adic integers, and back, from path to index representation.

x = a^

It is convenient to rewrite **a** and **x** in form of scalar product.
Consider **N+1** element vectors $M_N$ and $P_N$

$M_N = (m_0, m_1, m_2, \ldots, m_N)$

where $0 \le m_j < P$, $0 \le j \le N$

$P_N = (P^0, P^1, P^2, \ldots, P^N)$

Then

$x = m_0 P^0 + m_1 P^1 + m_2 P^2 + \ldots + m_N P^N$

can be represented as scalar product of two vectors

$x = (M_N \bullet P_N^T)$

$^T$ – as usual, means operation of matrix transposition (i.e. changing rows to columns). While

$a = m_0 P^N + m_1 P^{N-1} + m_2 P^{N-2} + \ldots + m_N P^0$

is

$a = (M_N^R \bullet P_N^T) = (M_N \bullet (P_N^R)^T)$

Here **R** means reverting elements of a vector. It is obvious that operator **^** is idempotent

x^^ = a^ = x

The important thing about this trivial operation is that we can perform arithmetic operation on points of a grid, and then immediately find a path to it by applying operator **^**, and vice versa, for a given path we can find a corresponding grid point.

It is convenient to define operator **[]** as coefficient in scalar representation:
Let x be as previously

$x = m_0 P^0 + m_1 P^1 + m_2 P^2 + \ldots + m_N P^N$

Then

$x[i] = m_i$

It is easy to see that

$x[i] = x^{\wedge}[N-i]$

## Mapping subintervals to paths

Now any subinterval can be mapped to a pair of paths on a coding tree, provided that edge points of subintervals belong to some $G(P^N)$. We will use notation [ , ] for intervals presented as a pair of indexes, where $l \leq r$ and $[l^{\wedge}, r^{\wedge}]$ – as a pair of paths and [ , ) for pairs paths to subintervals. A simple fact, just to note:

if an interval $[l, r]$ lies inside an interval $[l_1, r_1]$, then $d_P(l_1^{\wedge}, r_1^{\wedge}) \leq d_P(l^{\wedge}, r^{\wedge})$

In other words, paths to subinterval's edges are closer then paths to enveloping interval. So, if an interval's edges have a common path, then paths to edges of any subinterval have at least the same or even longer common paths. A length of common path can be calculated as $\text{ord}_P(r^{\wedge} - l^{\wedge})$. As an example see *Picture 5*.

Let discuss in more details the rightmost semi interval, i.e. a subinterval which ends at point 1. This point has index equal to $P^N$. Because we are working in the ring on integers numbers mod $P^N$ the index is equal to 0 in this ring. So path to 1 has the form {0, 0, … , 0}, an general form of the rightmost interval is $[l, 0]$.

What is a p-adic length of a rightmost interval? By definition

$d_P(l,0) = | 0 - l |_P = | -l |_P$

And common length is $\text{ord}_P(-l)$.

What is a "negative" path in our case? Path is always a path to some point on a grid. We use indexes for representing them. So a negative path may be defined as a path to a point, represented by negated index.

$-l = (-(l^{\wedge}))^{\wedge}$

By definition negative numbers in ring mod $P^N$ are

$l^{\wedge} + -(l^{\wedge}) = 0 \quad \text{mod } P^N$

## Common paths

Consider common part of all paths to points of semi interval $[l, r)$ ($l$ and $r$ are indexes here); both of them belongs to $G(P^N)$. All these paths end at corresponding points $p_i$

$l \leq p_i \leq r-1$

(remember that $r$ does not belong to subinterval). What is a common path to all these point?
First consider length of a common path. To find it we may first find maximum p-adic distance among all pairs:

$\max(|p^{\wedge} - q^{\wedge}|_P) ; \quad l \leq p < q \leq r-1$

We can use ultrametric feature of p-adic norm (see, for example, [**Koblitz, Holly**]):

$|x-y|_P <= \max(|x|_P, |y|_P)$

In our case we can use it as

$(|p^{\wedge} - q^{\wedge}|_P) = |(p^{\wedge} - l^{\wedge}) + (l^{\wedge} - q^{\wedge})|_P \leq \max(|(p^{\wedge} - l^{\wedge})|_P , |(l^{\wedge} - q^{\wedge})|_P)$

$l \leq p < q \leq r-1$

So all we need is to find

$\max(|(p_i^{\wedge} - l^{\wedge})|_P) ; \quad l < p_i \leq r-1$

which, by construction, is:

|(r -1)^ - l^|$_P$

Now we can calculate length of a common path. Special case l = r – 1 is important but trivial – the length here is simply a length of l. If l ≠ r – 1 then it is equal to ord$_P$((r-1)^ - l^). Because function ord$_P$ is not defined for zero argument we introduce a function com, defined on p-adic numbers:

com$_{P,N}$(l, r) = N if ( l == r ) else ord$_P$(r - l)

If l, r belongs to G(P$^N$) then length of common path is calculated as com$_{P,N}$(l^, r^).

Finally we have:

Paths to points of semi interval [l, r) have a common path of length com$_{P,N}$(l^, (r-1)^).

Common path is a sub path of length com$_{P,N}$(^, (r-1)^) of l^ starting from root.

In the following table examples different intervals of level 2 from *Picture 6* and their common paths are shown.

| l^ | r^ | l | r | r-1 | l | (r-1) - l | com$_{2,2}$( l^, r^ ) | Common path |
|---|---|---|---|---|---|---|---|---|
| {0, 0} | {0, 1} | 0 | 1 | 0 | 0 | 0 | 2 | {0, 0} |
| {0, 0} | {1, 0} | 0 | 2 | 1 | 0 | 2 | 1 | {0} |
| {0, 0} | {1, 1} | 0 | 3 | 2 | 0 | 1 | 0 | { } |
| {0, 0} | {0, 0} | 0 | 0 | 3 | 0 | 3 | 0 | { } |
| {0, 1} | {1, 0} | 1 | 2 | 1 | 2 | 0 | 2 | {0, 1} |
| {0, 1} | {1, 1} | 1 | 3 | 2 | 2 | 3 | 0 | { } |
| {0, 1} | {0, 0} | 1 | 0 | 3 | 2 | 1 | 0 | { } |
| {1, 0 } | {1, 1} | 2 | 3 | 2 | 1 | 1 | 2 | {1, 0} |
| {1, 0} | {0, 0} | 2 | 0 | 3 | 1 | 2 | 1 | {1} |
| {1, 1} | {0, 0} | 3 | 0 | 3 | 3 | 0 | 2 | {1, 1} |

## Rescaling based on P-adic distance (*PR*)

Consider two paths x = {0, 0, 1} and y = {0, 1, 1} or, as p-adic numbers:

x = 0*2$^0$ + 0*2$^1$ + 1*2$^2$ = 4

y = 0*2$^0$ + 1*2$^1$ + 1*2$^2$ = 6

or, as grid points

x^*2$^{-3}$ = (1*2$^0$ + 0*2$^1$ + 0*2$^2$)/8 = 1/8

y^*2$^{-3}$ = (1*2$^0$ + 1*2$^1$ + 0*2$^2$)/8 = 3/8

Because all subintervals in coding process will be inside [1/8, 3/8) all subsequent intervals will be inside it (this is how coding works), that means that all paths to these subintervals will have a common part. We can calculate common path of x and y according to the procedure described above:

y^ -1 = 1*2$^0$ + 1*2$^1$ + 0*2$^2$ = 6

(y^ -1)^ = 0*2$^0$ + 1*2$^1$ + 0*2$^2$ = 2

(y^ -1)^ - x = 2

And finally

com$_{2,3}$(x, y) = ord$_2$(2) = 1

We can store this path as a vector of coefficients and proceed with remaining part.

To make further descriptions shorter we introduce two operators: extracting and rescaling
Extracting is a trivial operator - it creates a vector of the first j coefficients of x:

$$x = m_0 P^0 + m_1 P^1 + m_2 P^2 + \ldots + m_n P^n$$

$$\text{ext}(x, j) = \{m_0, m_1, m_2, \ldots, m_{j-1}\}$$

if second argument is omitted, then all coefficients are extracted:

$$\text{ext}(x) = \{m_0, m_1, m_2, \ldots, m_n\}$$

One more operation on vector representation:

$$\text{cut}(x, n, m)$$

removes $m$ bit starting with position $n$ and shrink the vector of coefficients.
Rescaling is just omitting first $j$ terms in $x$ and removing common factor $P^j$,

$$\text{res}(x, j) = m_j P^0 + m_{j+i} P^1 + m_{j+2} P^2 + \ldots + m_n P^{n-j}$$

Why do we call it rescaling? Because we can continue with level $n-j$ as a first level and do not care about previous steps.
Let see what happens with corresponding index

$$\text{res}(x, j)\wedge = m_j P^{n-j} + m_{j+i} P^{n-j+1} + m_{j+2} P^{n-j+2} + \ldots + m_n P^0$$

As an integer number it is smaller then the original one. Rescaling keeps numbers from growing and makes it possible to use computer's integer arithmetic (not infinite precision) for calculations, which makes this algorithm robust.
Continuing our example (do not forget remove common factor!)

$$\text{res}(x,1) = (0*2^1 + 1*2^2) / 2 = 0*2^0 + 1*2^1$$

$$\text{res}(y,1) = (1*2^1 + 1*2^2) / 2 = 1*2^0 + 1*2^1$$

Indexes will be

$$\text{res}(x,1)\wedge = 1$$

$$\text{res}(y,1)\wedge = 3$$

and grid points

$$\text{res}(x,1)\wedge *2^{-2} = 1/4$$

$$\text{res}(y,1)\wedge *2^{-2} = 3/4$$

We can show how this works for weight interval

{c:[0, 0.125), b:[ 0.125, 0.375), d:[0.375, 05), a:[0.5, 1)}

and message {b}.

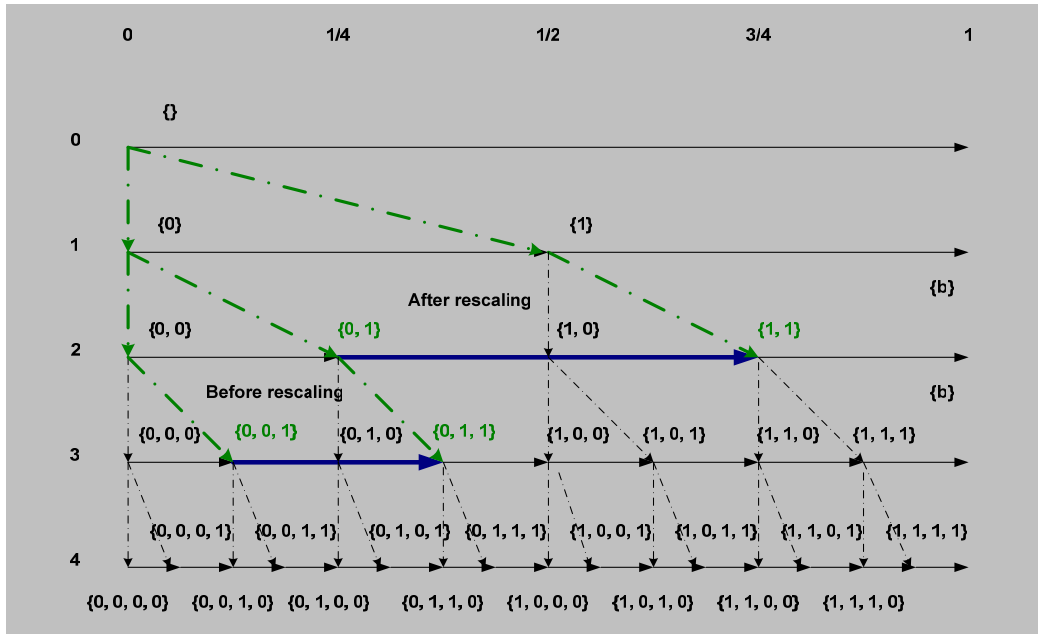

*Picture 7. Rescaling*

In arithmetic coding analogous procedure (see [**Bodden**]) is called *E1/E2*. We will call this rescaling *PR* (**p**-adic **r**escaling).

Trivial, but important case is when an interval occupies a whole subinterval of level K. In this case $x\char"5E = y\char"5E - 1$ and the interval can be rescale on full length to starting interval [0, 1). This fact will be used later in discussion how p-adic coding corresponds to Huffman algorithm.

## Lifting

In all our previous considerations and examples we use grids of minimal level. We may as well fix a level deep enough to perform all calculations. In fact, adding or removing trailing zeros in path representation does not change p-adic representation of a point, but, of course, changes its index representation. We will call operation of adding or removing zeros *lifting*. The reason for this name is that on a picture it looks like moving points in vertical direction.

There are several advantages of using fix level in calculations:
- It may be easy and more efficient coded, especially for P=2.
- Model may be unable to present results as numbers of current ring $G(P^N)$; in this case special procedure must be implemented for changing level on a model's demand.

Let x to be from $G(P^N)$. Lifting is a mapping x to $G(P^{N+f})$

$x = m_0 P^0 + m_1 P^1 + m_2 P^2 + \ldots + m_N P^N$

lift(x, j) => $x = m_0 P^0 + m_1 P^1 + m_2 P^2 + \ldots + m_N P^N + 0 \cdot P^{N+1} + \ldots + 0 \cdot P^{N+j}$

where j ≥ 0

Evidently as an integer number x does not change, but as an index it changes dramatically. Important, but trivial feature of lifting is that it does not change common paths.
Lifting can be defined also for negative argument:

lift(x, -j) => $x = m_0 P^0 + m_1 P^1 + m_2 P^2 + \ldots + m_{N-j} P^{N-j}$

where j ≥ 0

If last j coefficient were zero, negative lifting also does not change x as an integer number. To use negative lifting without changing results we need to know an order of the last non zero coefficient:

lnz(x) = min( j: $m_k$ =0 for k>j)

This function gives us the highest possible for x level. A semi interval [x y) may be positioned at level

hpl(x,y) = max(lnz(x), lnz(y))

Our procedure for calculating common path length of a semi interval was defined for intervals at hpl level. To extend it for the case when an interval belongs to a fixed level we need first to lift it back to hpl.

$com_{P,N}$( x, y ) = N if ( x == y ) else $ord_P$(x - y)

$com_{P,N}$( x, y ) = $com_{P,hpl(x,y)}$ (lift(x, hpl(x,y) - N ), lift(y, hpl(x,y) -N ) )

Fortunately we do not have to go in that complication. The reason for this is that lifting does not change number of common paths.

Let explore previous example restricting all calculations to level 4.

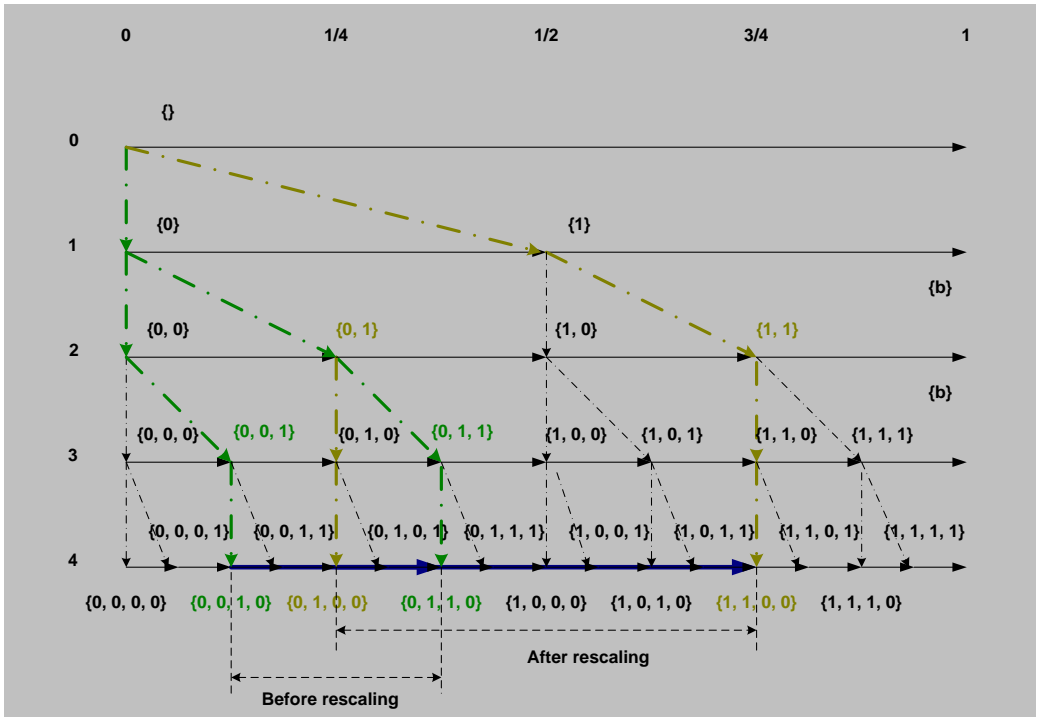

*Picture 7a. Rescaling on level 4*

Consider two paths x and y from the previous example, but fixed the level equal to 4. On this level x and y can be presented as {0, 0, 1, 0} and {0, 1, 1, 0} or, as p-adic numbers:

x = $0*2^0 + 0*2^1 + 1*2^2 + 0*2^3$ = 4

y = $0*2^0 + 1*2^1 + 1*2^2 + 0*2^3$ = 6

To determine common path length we need to calculate

y^ -1 = 5

(y^ -1)^ = 10

(y^ -1)^ - x = 6

And finally

com$_{2,4}$(x, y) = ord$_2$(6) = 1

After rescaling we have new x and y:

x = 0*2$^0$ + 1*2$^1$ + 0*2$^2$ = 2

y = 1*2$^0$ + 1*2$^1$ + 0*2$^2$ = 3

But they belong to level 3. To return x and y back to level 4 lifting is needed:

lift(x, 1) = 0*2$^0$ + 1*2$^1$ + 0*2$^2$ + 0*2$^3$

lift(y, 1) = 1*2$^0$ + 1*2$^1$ + 0*2$^2$ + 0*2$^3$

## Finding the shortest path point

When coding is over we can choose any paths to any point from a final semi interval as a result. But points from the same semi interval may and have different paths after dropping trailing zeros. Let take a simple example when a message finally ends with semi interval [g$_5$(4), g$_{10}$(4)). Because we can drop trailing zeros, point g$_8$(4) is the best choice – after dropping trailing zeros it becomes g$_1$(1).

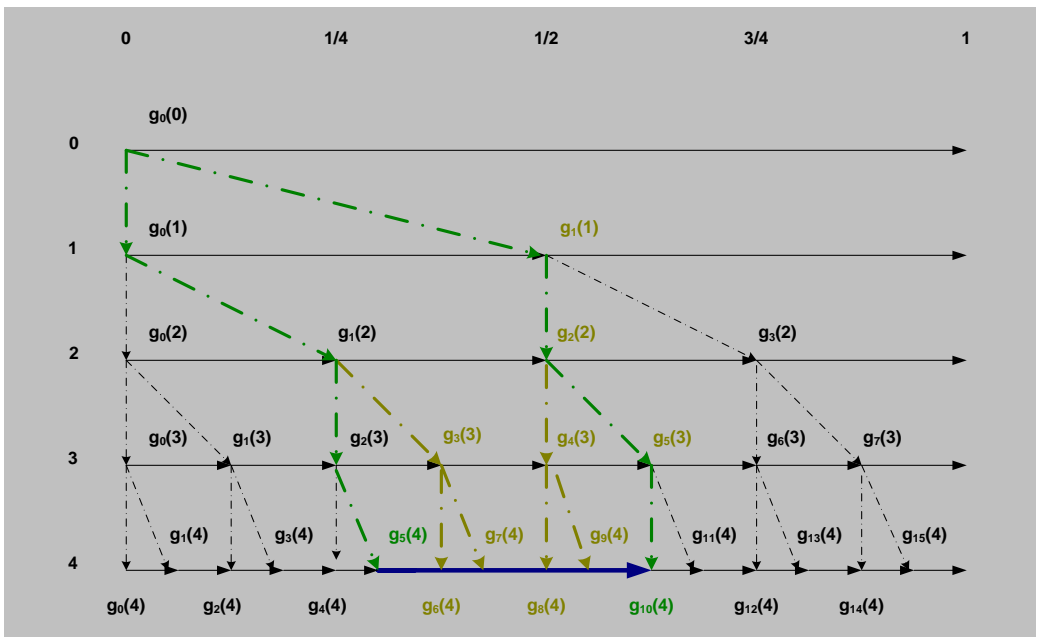

*Picture 8. Shortest path point*

A shortest path point in a semi interval [l, r) can be defined as a point with minimum level.

lv(x^) = max(i: m$_i$ ≠ 0)

g = min(lv(x^): l ≤ x < r )

But let consider paths as integers. From this point of view a point with minimal path is just a minimal p-adic integer. So

g = min(x^: l ≤ x < r)

We can check this for our example:

| Point | g$_5$(4) | g$_6$(4) | g$_7$(4) | g$_8$(4) | g$_9$(4) | g$_{10}$(4) |
|---|---|---|---|---|---|---|
| Path | {0, 1, 0, 1} | {0, 1, 1, 0} | {0, 1, 1, 1} | {1, 0, 0, 0} | {1, 0, 0, 1} | {1, 0, 1, 0} |
| p-adic number | 10 | 6 | 15 | 1 | 9 | 5 |
| Index | 5 | 6 | 7 | 8 | 9 | 10 |

## *Model*

Model is just an abstraction for a set of functions. One function calculates new subinterval on a base of incoming symbol and current interval in a predefined grid.

M.code(a, l, r) => $l_{new}$, $r_{new}$

An other takes as arguments a point and current interval and returns a new subinterval and a symbol

M.decode(g, l, r ) => $l_{new}$, $r_{new}$, a

l, r, g belongs to grid G($P^N$), a – to alphabet A.
Model operates with indexes from a ring of nonnegative integers modular $P^N$, so we have three possible variants how one subinterval on a ring can be situated inside another:

$0 \le l \le l_{new} < r_{new} \le r < P^N$

r=0 ; $0 \le l \le l_{new} < r_{new} < P^N$

r=0 ; $r_{new} = 0$; $0 \le l \le l_{new} < P^N$

And, of course some technical things: initialization and taking care of end of message.

M.init(A, P, N, x)

Where A is alphabet, P, N – characteristics of grid G($P^N$), x – optional parameter, some auxiliary information, which may be used by a model for optimization.
code and decode functions may update model, but they must do it in sync.

## *Input and Output*

I (*input*) and O (*Output*) are abstracts for pushing and receiving information.
To make notations short we introduce an ugly term P-bit, which means one of symbols 0, … P-1. For P=2 it is obviously a normal bit. Now let describe input and output operations.

I.getC => returns next character from input stream or EOM (*End Of Message*)

I.getB(n) => returns  next n P-bit vector from input stream

O.pushB(U ) – pushes all P-bits from vector U

O.pushB(p, n) – pushes P-bit p n times

O.pushC(a) – pushes a symbol to an output stream

## *Algorithms*

Now we are in position to describe the p-adic coding algorithm. Main idea of this algorithm is the same as in arithmetic coding – a message is mapped to in interval on [0, 1).

There two parts of the algorithm – encoding and decoding, but whatever we are doing the first step – initialize a model:

M.init(A, N)

### Coding

Start with an empty message – no symbols. An empty message is coded as [0, 1), empty path U = {} or as [0, 0).

l, r = 0, 0

When a symbol a comes

a = I.getC

model calculates a new interval.

l, r = M.code(a, l, r)

Now calculate a common path length

n = com$_{P,N}$(l^, (r-1)^)

If n > 0 we can push common path to an output

O.pushB(ext (l^, n))

and do rescaling.

l^, r^ = res(l^, n), res(r^, n)

And we also need to lift rescaled values back to level N and convert to index representation.

l, r = lift(l^, n)^, lift(r^, n)^

Now we can read a next symbol and repeat steps.

## Pseudo code

```
M.init(A, P, N)
l, r = 0, 0
while ( ( a = I.getC ) != EOM ) {
    l, r = M.code(a, l, r)
    n = com_P,N(l^, (r-1)^)
     if ( n > 0 ) {
         O.pushB(ext(l^, n))
         l, r = lift(res(l^, n), n)^, lift(res(r^, n), n)^
     } //if
} //while
l, r = M.code(EOM, l, r)
q = selectPoint(l, r)
O.pushB(ext(q, lnz(q))
```

We do not specify here what selectPoint does. The only requirement is to return a grid point from final semi interval [l, r), but of cause, it's a good idea to return a point with a shortest paths. As it follows from previous discussion, all we need is to find a minimal integer in p-adic representation. So, to select a point with minimal path we should define

selectPoint( l, r ) = min(x^: l ≤ x < r)

lnz used here not to push trailing zeros.

## Decoding

Start with an empty message – no symbols. An empty message is coded as [0, 1), or as empty path U={} or as pair of indexes:

l, r = 0, 0

As the first step read first N P-bits from an input stream and construct a number from the vector. We need also to transform a path we a getting from a stream, to a number, so we use operator ^.

g= (I.getB(n) • P$^T_N$)^

where P$^T_N$ is a vector

P$_N$ = (P$^0$, P$^1$, P$^2$, … , P$^N$)

Model calculates a new interval and a symbol a

M.decode(g, l, r ) => l, r, a

Now, a is a new decoded symbol and can be pushed into a stream of decoded symbols

O.pushC(a)

Next, as in the coding algorithm, calculate common path length

$n = com_{P,N}(l\hat{}, (r-1)\hat{})$

If $n > 0$ we can drop common path and do rescaling.

$l, r, g = lift(res(l\hat{}, n), n)\hat{}, lift(res(r\hat{}, n), n)\hat{}, lift(res(g\hat{}, n))\hat{}$

read additional n P-bits and recalculate g

$g = g + (I.getB(n) \cdot P^T_n)\hat{}$

Now we can repeat all steps.

**Pseudo code**

```
M.init(A, P, N)
l, r = 0, 0
g = (I.getB(n) • P^T_N)^
while ( true ) {
   l, r, a = M.decode(g, l, r )
   if ( a == EOM ) break
   O.pushC(a)
   n = com_P,N(l^, (r-1)^)
   if ( n > 0 ) {
       l, r, g = lift(res(l^, n), n)^, lift(res(r^, n), n)^, lift(res(g^, n),n)^
       g = g + (I.getB(n) • P^T_n)^
   } //if
 } //while
```

## *One important particular case – Huffman codes*

Now we are prepared to show that p-adic coding algorithm gives exactly the same codes as Huffman's algorithm [**Huffman**] if a weight interval is prepared in a special way. Let as assume that for a given alphabet and symbol probabilities a Huffman code tree was constructed. For example:

| Symbol (s): | a | b | C | d | e |
|---|---|---|---|---|---|
| Codeword (h(s)): | 000 | 001 | 10 | 01 | 11 |
| Grid level (cl(s)): | 3 | 3 | 2 | 2 | 2 |
| Starting index in grid: | 0 | 1 | 2 | 1 | 3 |

We can map the tree to weight interval using the same technique as we used for coding messages

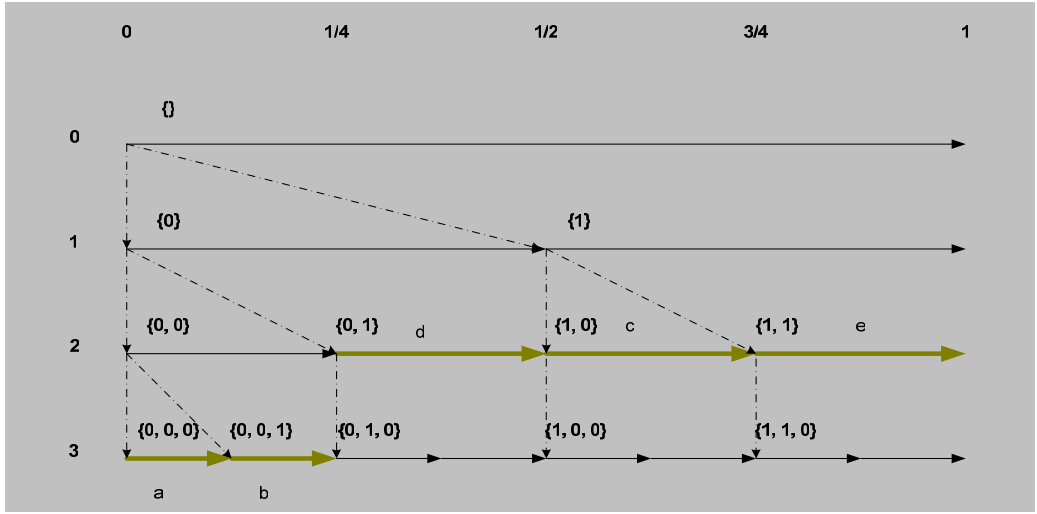

*Picture 9. Mapping Huffman code tree to weight interval*

After lifting all intervals to highest grid ($N=3$):

| Symbol (s): | a | B | c | d | e |
|---|---|---|---|---|---|
| Codeword (lift(h(s),N-cl(s))): | 000 | 001 | 100 | 010 | 110 |
| Starting index in grid: | 0 | 1 | 4 | 2 | 6 |

Algorithm of constructing weight intervals from a Huffman code tree for alphabet A is simple.

> Let cl(s) be a length of Huffman code of symbol s and $N = \max(cl(s))$ among all s from A, h(s) - Huffman code of s, then symbols s occupies a semi interval starting at point with index lift(h(s),N-cl(s))^ and ending at starting point of a next symbol or 1.

Constructed weight interval has an important property – all of subintervals occupy a whole grid interval of some level. It was shown above that in this situation left end right ends have an entire path in common; so *PR* rescaling will push all of it into an output and a next symbol will be coded starting with [0,1) interval. This proves that for this particular choice of weight interval p-adic coding works identical to Huffman's algorithm.

## *Another particular case – Golomb-Rice codes*

Surprisingly enough, but p-adic coding algorithm produces Golomb-Rice [**Golomb, Rice**] codes when supplied with single symbol alphabet and special model; no changes to algorithm itself are needed.

If an alphabet contains only one symbol, the only information a message may contains is its length. So coding of a message is equivalent to coding of a natural number – the length. We will use symbol * to identify the only entry.

The model is trivial:

M.code(*, l, r) => l, r-1

M.code(EOM, l, r) => r-1, r

M.decode(g, l, r ) => if (g == r-1) then l, r, EOM  else l, r-1, *

The algorithm will do all the work. Let start with $P=2$ and consider a grid $2^{N+1}$.
Coding procedure starts with

l = r = 0

If a message is empty, we have to encode EOM. To do this we need to calculate r-1= 0 - 1, which is $2^{N+1}-1$ and return a path to $2^{N+1}-1$. This path consists of N+1 ones: {1, 1, … , 1, 1}. This is our new representation of zero. If a symbol comes, the model recalculates r and l:

l = 0

r = $2^{N+1}-1$

If it was the only symbol in a message, then the model returns $2^{N+1}-2$, $2^{N+1}-1$, and a code is a path to a point with index $2^{N+1}-2$: {1, 1, … , 1, 0}. This procedure may be continued until a message's length is less than $2^N$. At this point the model returns

l = 0

r = $2^N$

because $com_{P,N}(0^\wedge, (2^N -1)^\wedge) = 1$ *PR* rescaling will be used; one 0 will be pushed to output buffer, l and r return to their initial values l = r = 0. The coder is in initial state and ready to receive a new symbol.

Encoder stays almost without changes. We have defined

selectPoint( l, r ) = l

And drop lnz call in the last pushB operation to keep trailing zeros

O.pushB(ext(q))

If a messages of length W comes $W/2^N$ zeros will be pushed in output buffer; the rest part of the output will contain a path to a point which index is $0 – (W\%2^N)$. After encoding EOM we have to move the point one step to the left. So finally index will be $0 – ((W\%2^N) + 1)$.

For example, for N=3 we have:

| W | code | W  | Code  |
|---|------|----|-------|
| 0 | 1111 | 8  | 01111 |
| 1 | 1110 | 9  | 01110 |
| 2 | 1101 | 10 | 01101 |
| 3 | 1100 | 11 | 01100 |
| 4 | 1011 | 12 | 01011 |
| 5 | 1010 | 13 | 01010 |
| 6 | 1001 | 14 | 01001 |
| 7 | 1000 | 15 | 01000 |

The codes look very much like Golomb-Rice codes. Indeed, they may be transformed to each other by replacing 1 with 0, and 0 with 1 - binary NOT.

There is no magic in changing unary representation and delimiter – there is no difference between counting a number 0 of before first 1 and counting number of 1 before first 0. Transformation of the rest part – after delimiter, may be not that clear.

In the ring of integers modular $2^N$

$0 – (R +1) = (2^N - 1) - R$

here $R = (W\%2^N)$; $R < 2^N$. In binary representation $(2^N – 1)$ is a vector U of N 1. Now

NOT(U – R) = R

This proves that after NOT transformation the rightmost part of codes transforms to $W\%2^N$.

Any prime P can be used with this model. But this generalization does not look very promising. In fact, the reason why we discuss Huffman and Golomb-Rice codes here is to emphasize that the most popular entropy codes have a common base – they all maps messages to p-adic integer numbers.

## *Rescaling based on Archimedean distance (AR)*

We were very ingenious when selecting most convenient for us weigh interval:

{a:[0, 0.5), b:[0.5, 0.75), c:[0.75, 0.875), d:[0.875,1)}

Yes, compression rate does not depend on an order of subintervals, but calculation and resulted codes do. Let shuffle the weigh interval:

{b:[0, 0.25), a:[0.25, 0.75), c:[0.75, 0.875), d:[0.875,1)}

Now subinterval a:[0.25, 0.75) covers the center point 1/2. Consider now a message containing only symbols a. It can be easily shown that left edge of message interval will be always less than 1/2, while the right one – greater. From p-adic point of view this means that $ord_p(l, r)$ is always zero and there is no common path and, as a sequence, rescaling will never happen. If we continue coding {a, a, … , a} we will end in integer overflow error or will be faced to use infinite precision arithmetic.

To save our integer arithmetic from huge numbers we have to use the fact that Archimedean length in this case is less or equal to 1/2.

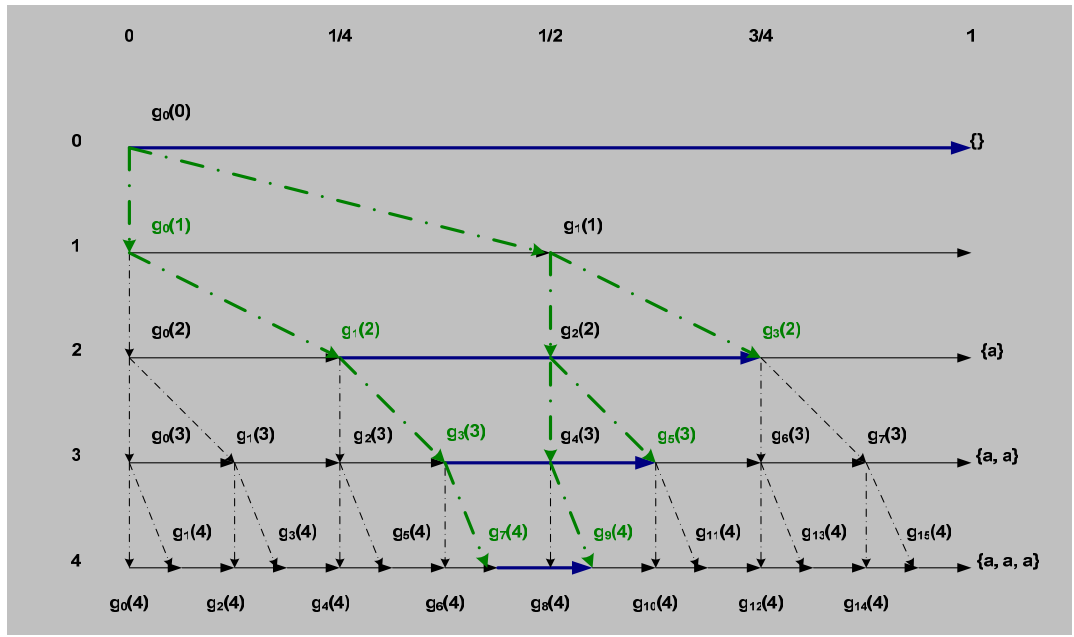

*Picture 10. Coding {a, a, a}*

For P ≠ 2 situation is more complex. A semi interval can include any grid point 0 < n < P. In the following example (P = 3) an interval has Archimedean length 2/9, but p-adic length 1.

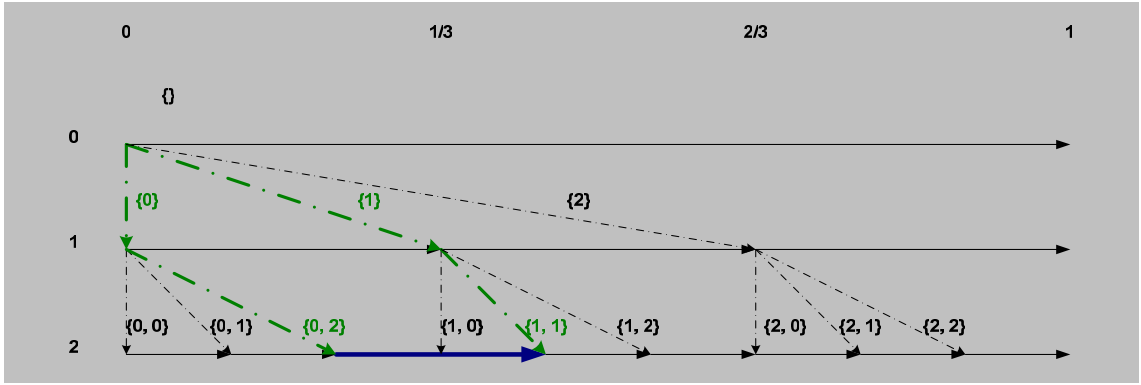

*Picture 11. Before rescaling*

Now let explore a case when a sub interval lies in the smallest interval of level 2, which includes a point of level 1 with index n. p-adic representation of left l and right r edges of such subinterval is.

l = {n-1, P-1, …. }

r = {n, 0, … }

It's Archimedean length is less or equal to 2/(P*P). We want to map it to a bigger interval, precisely to interval [n-1, n+1) from level 1. This can be done by a linear transformation:

$Y(X) = XP^1 - nP^0 + nP^{-1}$

Let's consider how a semi interval defined in p-adic representation as

$l = m_0 P^0 + m_1 P^1 + m_2 P^2 + \ldots + m_N P^N$

$r = k_0 P^0 + k_1 P^1 + k_2 P^2 + \ldots + k_N P^N$

transforms under this mapping. The first thing we need to do – to transform paths to points. We can do it by using **IP** transformation:

$a = m_0 P^{-1} + m_1 P^{-2} + m_2 P^{-3} + \ldots + m_N P^{-N-1}$

$b = k_0 P^{-1} + k_1 P^{-2} + k_2 P^{-3} + \ldots + k_N P^{-N-1}$

Now we can apply linear transformation:

$Y(a) = (m_0 - n)P^0 + (m_1 + n)P^{-1} + m_2 P^{-2} + \ldots + m_N P^{-N}$

$Y(b) = (k_0 - n)P^0 + (k_1 + n)P^{-1} + k_2 P^{-2} + \ldots + k_N P^{-N}$

For this subinterval we have:

$m_0 = n - 1; \quad m_1 = P - 1$

$k_0 = n; \quad k_1 = 0$

So

$Y(a) = 0P^0 + (n - 1)P^{-1} + m_2 P^{-2} + \ldots + m_N P^{-N}$

$Y(b) = 0P^0 + nP^{-1} + k_2 P^{-2} + \ldots + k_N P^{-N}$

Rescaling will drop first zero terms. Reverting back from points to paths we can find how this transformation works on paths:

$Y(l) = (n - 1) P^0 + m_2 P^1 + \ldots + m_n P^{N-1}$

$Y(r) = nP^0 + k_2 P^1 + \ldots + k_n P^{N-1}$

Or in vector representation:

$Y(l) = \{n-1, P-1, ....\} \Rightarrow \{n-1, ...\}$

$Y(r) = \{n, 0, ...\} \Rightarrow \{n, ...\}$

we just remove second (counting from the left) elements.
It is also easy to verify that center point $\{n, 0, 0, ..., 0\}$ of this mapping is a stable point, i.e. $Y$ maps it to itself

$Y : \{n, 0, 0, ..., 0\} \Rightarrow \{n, 0, ..., 0\}$

New interval [l, r) contains the stable point.
Coming back to the example (here n=1) we can draw the picture after rescaling:

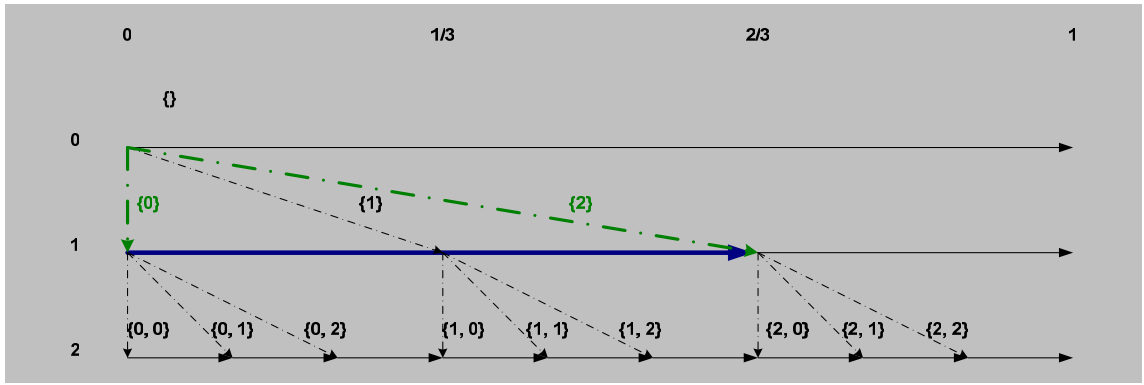

*Picture 11a. After rescaling*

We will refer this rescaling as **AR**. Important difference between **AR** and **PR** rescaling is that **AR** does not push anything in output buffer.

It is convenient to invent a special predicate **AR?** for testing if **AR** rescaling can be applied for an interval.

$AR?(l, r, P) = (r[0] - l[0] == 1)$ AND $(l[1] == P-1)$ AND $(r[1] == 0)$

To continue coding we must remember the applied mapping, it can be done by storing only two parameters:
n – a stable point and u – a number of times rescaling was applied.

What may happen if we continue coding?
1. [l, r) are still contains n
    1.1. value of AR? predicate is false
    1.2. value of AR? predicate is true
2. [l, r) does not contain n
    2.1. n lays to the right of r;  toRight?( n, r) == true
    2.2. n lays to the left of l;    toLeft?( n, l) == true

To test condition 2.1 and 2.2 we introduced two predicates *toRight?* and *toLetf?*. There predicates are suppose to receive a path as second argument, i.e. a number in *p*-adic integer number; first argument is an integer number

$toRight?(n, r) = (r[0] < n)$ OR $(r\char`\^ == n)$

$toLeft?(n, l) = l[0] \geq n$

Now let discuss situations mentioned above:

1.1. This is the simplest case. We just continue coding.

1.2. Increase u: u = u +1; do **AR** rescaling and continue coding.

2.1. This means that the whole interval lays in [{n-1, P-1, ... , P-1}, {n, 0, ... , 0}); where P-1 is added u times. Any subinterval from this interval has common path {n-1, P-1, ... , P-1}, so we can now push this path into output and rescale l and r one more time removing fist digits.

2.2. This means that the whole interval lays in [{n, 0, ... , 0, 1}, { n, 0, ... , 0,1}); where 0 is added u times. Any subinterval from this interval has common path {n, 0, ... , 0}, so we can now push this path into output and rescale l and r one more time removing fist digits.

*AR* and *PR* rescaling procedures together guaranty that current coding interval will never be smaller than $2/P^2 - 1/P^N$. This means that maximum value of indexes is $2P^{N-2} - 1$.

## *Algorithms revised*

### Coding with *AR*

A new feature here, comparing to the first variant of p-adic encoding algorithm, is that we need to track *AR* transformation. To do this we introduce two new variables sp and spn.
- sp – stable point of *AR*; it is a point of level 1 and may be represented as a positive integer (not path) 0 < sp < P.
- spn – number of times *AR* was applied.

Some additional operations should be done at final step. First of all we need to check, as in the main loop, if the final interval is situated to the left or to the right of a stable point and, if this is the case, do necessary pushing and then proceed to usual final search for minimal point. If not and spn is not zero, then we are lucky and we already have a point from level 1 and all we need to do is just to push out sp.

### Pseudo code

```
M.init(A, N)
l, r = 0, 0
sp, spn = 0, 0

while ( ( a = I.getC ) != EOM ) {
   l, r = M.code(a, l, r)

   if ( spn ≠ 0 ) {
      if ( toLeft?(sp, l^) ) {
         O.pushB(sp, 1)
         O.pushB(0,spn)
         l, r = lift(res(l^, 1), 1)^, lift(res(r^, 1), 1)^
         sp, spn = 0, 0
      } //if
      if ( toRight?(sp, r^) ) {
         O.pushB(sp - 1, 1)
         O.pushB(P - 1, spn)
         l, r = lift(res(l^, 1), 1)^, lift(res(r^, 1), 1)^
         sp, spn = 0, 0
      } //if
   } //if

   // PR rescaling
   if ( spn == 0 ) {
      n = com_{P,N}(l^, (r - 1)^)
      if ( n > 0 ) {
          O.pushB(ext(l^, n))
          l, r = lift(res(l^, n), n)^, lift(res(r^, n), n)^
```

```
        } //if
    }
    //AR rescaling
    while ( AR?(l^, r^) ) {
        sp =  r^[0]   if sp == 0
        spn = spn + 1
        l, r = lift(cut(l^,1,1),1)^, lift(cut(r^,1,1),1)^
    } //while

} //while

l, r = M.code(EOM, l, r)

if ( spn ≠ 0 ) {
   if ( toLeft?(sp, l^) ) {
      O.pushB(sp, 1)
      O.pushB(0, spn)
      l, r = lift(res(l^, 1), 1)^, lift(res(r^, 1), 1)^
      sp, spn = 0, 0
   } //if
   if ( toRight?(sp, r^) ) {
      O.pushB(sp - 1, 1)
      O.pushB(P - 1, spn)
      l, r = lift(res(l^, 1), 1)^, lift(res(r^, 1), 1)^
      sp, spn = 0, 0
   } //if
} //if
if (spn == 0) {
    q = selectPoint(l, r)
    O.pushB(ext(q, lnz(q))
} else {
    O.pushB(sp, 1)  // we already have point of level 1
} //if

//the End
```

## Decoding with *AR*

*AR* rescaling is simpler for decoding process, because we do not care about pushing anything out and a final step is most simple – we just finish decoding. The only thing which is new is additional reading from an input stream.

### Pseudo code

```
M.init(A, N)
l, r = 0, 0
spn = sp = 0
g = (I.getB(N) • P^T_N)^
while ( true ) {
   l, r, a = M.decode(g, l, r )
   if ( a == EOM ) break
   O.pushC(a)

   if ( spn ≠ 0 ) {
      if ( toLeft?(sp, l^) OR toRight?(sp, r^) ) {
         l, r, g = lift(res(l^, 1), 1)^, lift(res(r^, 1), 1)^ , lift(res(g^, 1), 1)^
         g = g + (I.getB(1) • P^T_1)^
```

```
          sp, spn = 0, 0
        } //if
     } //if

     // PR rescaling
     n = com_{P,N}(l^, (r-1)^)
      if ( n > 0 ) {
          l, r, g = lift(res(l^, n), n)^, lift(res(r^, n), n)^, lift(res(g^, n),n)^
          g = g + (I.getB(n) • P^T_n)^
        } //if

     // AR rescaling
     while ( AR?(l^, r^) ) {
         sp =  r^[0]  if sp == 0
         spn = spn +1
         l, r, g = lift(cut(l^,1,1),1)^, lift(cut(r^,1,1),1)^, lift(cut(g^,1,1),1)^
         g = g + (I.getB(1) • P^T_1)^
     } //while

 } //while

//the End
```

Of course, $P^T_1$ is just 1 and we can also omit ^ operator. The operation

g = g + (I.getB(1) • $P^T_1$)^

can be replaced (in two places) by

 g = g + I.getB(1)

## Implementation

We have implemented all algorithms and all tests in Ruby [**Ruby**] – a new popular interpreted, dynamically typed, pure object-oriented, scripting language. And Ruby proved to be very helpful. We would hardly be able to try so many variants and run innumerous tests in any other language.

## *P=2*

Now let discuss the practical case of P=2. All previous discussion remains valid – this is just a special case. This case has most important advantage – we can use real bits and binary vectors. This is extremely convenient.

All algorithms remain the same. Only some small improvement can be done for *AR* rescaling. Because the only possible value for n is 1, there is no need to store it as spt. In case when toLeft? returns true we have to push 1 and a number of 0; if toRight? returns true we have to push 0 and a number of 1.

## Arithmetic coding

We can see that arithmetic coding is just a special case of p-adic coding for P=2. All conditions expressed there as arithmetic operations can be done on bit level. In fact, many practical implementations use shifts instead.

Let us examine *E1* condition:

mHigh < g_Half

where

g_Half = 0x40000000

This condition means that most significant bit in binary representation of mHigh must be 0. This is also true for gLow because mLow < mHigh. Reverting to paths we can see that both gLow and gHigh have most significant bits in p-adic representation are equal to 0, so p-adic distance is less than 1 and *PR* condition is fulfilled.

However, in p-adic coding algorithm *PR* rescaling works for mHigh equal to g_Half. It is this small difference makes p-adic coding algorithm works exactly as Huffman algorithm for certain models. Arithmetic coding in this situation does not provide optimal compression (see discussion in [**Bodden**]).

*AR* rescaling is similar to *E3*. AR? predicate is equivalent to

 (g_FisrtQuater <= mlow) AND (mHigh < g_ThirdQuater)

We have implemented the same model as proposed in [**Bodden**] and get the same compression for all standard tests.

## *Results*

### Standard tests

For testing we used Calgary/Canterbury text compression corpus – popular set of tests first discussed in [**Bell**]. It contains files *bib, book1, book2, geo, news, obj1, obj2, paper1, paper2, paper3, paper4, paper5, paper6, pic, progc, progl, progp* and *trans*. These files may be obtained from [**Canterbury corpus**].

### Comparison with Arithmetic coding

We used program codes published in [**Bodden**] to get results of arithmetic coding. This program adds additional 4 bytes to an output file; these 4 bytes are in most examples the only difference between arithmetic and p-adic coding results.
In our test we use p-adic coding with P=2 and N= 31.

## *Conclusion*

Tree is a well known and widely used data structure in computer science. Arithmetic, Huffman and Golomb-Rice coding are also well known and widely used for a long time algorithms. p-adic numbers, ultrametric spaces are not so popular in computer science; even for pure mathematic they are relatively new. Is there any connection between them? We hope that we have shown this connection and that this connection is quite natural and fundamental.

A message, as sequences of symbols, may be considered as path on a tree. There are numerous ways to construct this mapping. It is quite fundamental and widely used way for presenting messages and is very popular in computer science and applications. On the other hand, trees are great models of p-adic numbers; many strange and unusual features of ultrametric spaces can be understood and visualized on trees [**Holly**]. This works also in the reverse direction – p-adic numbers is convenient tool for indexing paths and p-adic norm is a natural measure on trees.

On the other hand, a message can be mapped to a subinterval of a unit interval – this what real number arithmetic algorithm does. While theoretically clear and simple, this method was never used in practice, because of its inefficiency due to problems with computer based real arithmetic. Integer arithmetic coding solved this problem by introducing some practical receipts how to use integer numbers, instead of real ones. The resulting algorithm proved to be efficient and robust and may be because of this fact no theoretical analysis has been done. Integer version looks pretty much like the original algorithm, but in fact, difference between them is considerable; while real number algorithm works on a field, its integer number variant deals with a finite ring.

p-adic number coding algorithm explicitly works with numbers from the finite ring of positive integer numbers modular $P^N$. These numbers, being mapped to a union interval, create an equidistance grid $G(P^N)$. The next step is to create a path from a root through grid points of upper levels ($G(P^K)$; k=0..N-1) to points of this grid. This construction creates a bridge between ultrametric space of tree paths and Archimedean space of grid points. Now we can identify any grid point not only by its index, but by a path, i.e. by some p-adic integer number; the reverse is also true – any path can be identify by its end point from the grid and as so by an index. This dualism is the real base of p-adic arithmetic coding algorithm. We found a simple and elegant way to transform paths to indexes and back. We called it *IP* transformation. As a transformation from paths to points *IP* transformation can be considered as Kozyrev's transformation for finite paths, but *IP* transformation is reversible.

p-adic arithmetic coding algorithm works as a bridge between two spaces – ultrametric space of paths and Archimedean space of grid points. Model calculates intervals with edge points on the grid and then *IP* transformation maps them to paths; if these paths are closed to each other as p-adic numbers, then common path is pushed to output buffer, these p-adic numbers are truncated, and *IP* transformation maps them back on grid. For P=2 *PR* rescaling works pretty much like *E1/E2* rescaling but it has one small improvement. It is this improvement that makes it possible to show that for certain models and alphabet p-adic coding algorithm works as Huffman and Golomb-Rice algorithm. For P=2 and general models it works as arithmetic coding.

So we may say that three most popular entropy coding algorithms can be considered as special cases of one algorithm - p-adic coding, working with different P, models and alphabets. This also gives an answer to the question in the begging of this paragraph - arithmetic, Huffman and Golomb-Rice coding algorithms maps messages to ultrametric space of p-adic numbers. They are "*speaking in prose"*!